\documentclass[a4paper,12pt]{article}
\pdfoutput=1

\usepackage{amssymb}
\usepackage{amsmath}
\usepackage{hyperref}
\usepackage[utf8x]{inputenc}
\usepackage{geometry}
\usepackage{graphicx}

\usepackage[singlelinecheck=true]{subfig}
\usepackage{caption}
\usepackage{titlesec}
\titleformat{\section}
  {\normalfont\fontsize{15}{18}\bfseries}{\thesection}{1em}{}

\hypersetup{
    colorlinks=true,         
    linkcolor=blue,          
    citecolor=red,
    urlcolor=blue
}

\newcommand{\R}{\mathbb{R}}

\newcommand{\rmd}{\mathrm{d}}

\begin{document}

${}$\\
\begin{center}
\vspace{-16pt}
{ \large \bf Exploring Torus Universes\\ \vspace{10pt} 
in\\ \vspace{16pt} 
Causal Dynamical Triangulations}

\vspace{40pt}

{\sl T.G.\ Budd}$\,^{a}$
and {\sl R.\ Loll}$\,^{b,c}$

\vspace{24pt}
{\footnotesize

$^a$~The Niels Bohr Institute, Copenhagen University\\
Blegdamsvej 17, DK-2100 Copenhagen \O , Denmark.\\
{email: budd@nbi.dk}\\

\vspace{10pt}

$^b$~Radboud University Nijmegen,\\
Institute for Mathematics, Astrophysics and Particle Physics, \\
Heyendaalseweg 135, NL-6525 AJ Nijmegen, The Netherlands.\\
{email: r.loll@science.ru.nl}\\

\vspace{10pt}

$^c$~Perimeter Institute for Theoretical Physics,\\
31 Caroline St N, Waterloo, Ontario N2L 2Y5, Canada.\\
{email: rloll@perimeterinstitute.ca}\\

}

\vspace{36pt}


{\bf Abstract}
\end{center}

\noindent Motivated by the search for new observables in nonperturbative quantum gravity, 
we consider Causal Dynamical Triangulations (CDT) in 2+1 dimensions with the spatial topology of a torus.
This system is of particular interest, because one can study not only the global scale factor, but also 
global shape variables in the presence of arbitrary quantum fluctuations of the geometry. 
Our initial investigation focusses on the dynamics of the scale factor and uncovers a qualitatively new behaviour, 
which leads us to investigate a novel type of boundary conditions for the path integral. 
Comparing large-scale features of the emergent quantum geometry in numerical simulations  
with a classical minisuperspace formulation, we find partial agreement. By measuring the
correlation matrix of volume fluctuations we succeed in reconstructing the effective action for the scale factor 
directly from the simulation data. Apart from setting the stage for the analysis of shape dynamics on the
torus, the new set-up highlights the role of nontrivial boundaries and topology. 

\newpage

\section{Nonperturbative quantum gravity and observables}\label{sec:Introduction}

A central quest in any approach to nonperturbative quantum gravity is for the identification and evaluation of 
{\it observables}: finite, invariantly defined quantities characterizing ``quantum spacetime", the Planck-scale 
analogue of the curved spacetime described by the classical theory of General Relativity. One reason why such
observables are hard to come by is the a priori absence of a yardstick, in the form of
a preferred classical ``background geometry", with respect to which distances and volumes could be measured. 
In a full, nonperturbative quantum formulation, such a yardstick has to be extracted from the {\it dynamics} of the quantum gravity
theory, and therefore requires nontrivial knowledge of the latter. So far, only a few instances of such observables 
have been identified in specific candidate theories of 
quantum gravity. They typically depend on intrinsic properties of quantum geometry in a relational and nonlocal way.
Good examples of this are various notions of ``dimensionality", obtained by relating (suitable quantum analogues of) 
volumes to linear distances, say, in the form of a power law, which have been used for a long time in
quantum-gravitational theories obtained from dynamical triangulations \cite{dimensions}. 

The main motivation behind the work presented here is to try to push the quest for observables beyond what has been
considered so far. We will do this in a specific context, that of quantum gravity constructed from {\it Causal Dynamical
Triangulations (CDT)}. The CDT approach has made significant progress towards a nonperturbative realization of the 
path integral over higher-dimensional geometries in recent years, see \cite{physrep} for a comprehensive review. 
There is strong evidence 
that at large distance scales it describes a quantum universe fluctuating around an emergent classical background 
geometry. Its large-distance scaling properties, captured by the dynamically extracted Hausdorff \cite{ajl-prl,ajl-rec} and spectral 
\cite{spectral} dimensions, are 
compatible with those of a four-dimensional spacetime. Because of the nonperturbative
character of the quantum dynamics, there is no a priori guarantee that any emergent quantum spacetime will have
dimension four macroscopically. Checking that this happens is therefore an important part of verifying that a
quantum gravity theory possesses a good classical limit. 

In addition, by examining the expectation value of the 
spatial volume as a function of proper time, the spacetime emergent from CDT quantum gravity has been matched
with excellent accuracy to a de Sitter universe, including quantum fluctuations of the volume around it \cite{desitter}.
This ``volume profile" is another example of the type of geometric observable we are after. It is a quantity 
that can be accessed relatively easily by numerical measurements, and at the same time has a transparent
semi-classical geometric interpretation. From the point of view of General Relativity, it is simply the 
``scale factor" or ``Friedmann mode" of the classical metric field $g_{\mu\nu}(x)$. In cosmological 
approximations to the Einstein equations, where space is described as homogeneous and isotropic, the geometry
of spacetime is by assumption reduced to the dynamics of this global scale factor. In the context of full quantum gravity,
the quantum dynamics of this mode and its possible implications for cosmology are clearly important to
investigate, but they describe only one, global aspect of quantum spacetime. Can we go beyond the Friedmann mode
in understanding the structure of quantum geometry, and how?

A natural next step is to consider modes of the geometry which are still global, but describe {\it shape} rather than
scale. We will begin our exploration of this issue in spacetime dimension three. There are a number of
good reasons for doing so. Firstly, as far as large-scale geometric properties are concerned, CDT quantum gravity 
in 2+1 dimensions appears to resemble CDT quantum gravity in 3+1 dimensions; in fact, the analysis of
matching the
volume profile of the emergent quantum geometry with that of a de Sitter universe was first made in the former \cite{3dcdt}. 
Secondly, we have a neat and explicit way of isolating the conformal (``shape") degrees of freedom of 
the two-dimensional triangulated surfaces representing space in 2+1 dimensions, at least when
the spatial slices have torus topology \cite{modulidt,toappear}. By contrast, it does not appear
straightforward to isolate observables which are sensitive
to global shape in the full, four-dimensional quantum theory. Thirdly, the computational effort needed in the 
three-dimensional setting is significantly reduced compared to that in four dimensions. 

Caution is obviously advised when generalizing any results from three to four dimensions, since the local
dynamics and degrees of freedom of the two quantum gravity theories are expected to be very different. 
In part, this is already reflected
in their different phase structure (as statistical mechanical systems), which in four dimensions is much richer and has
recently been shown to contain a second-order phase transition \cite{phasetrans}, 
something not seen in three dimensions \cite{3dcdt}.
On the other hand, in evaluating the path integral of three-dimensional quantum gravity 
we will not make use of the possibility to reduce the dynamical (field) degrees of freedom to a finite number of physical
variables {\it before} the quantization, an essential step taken in most other treatments of quantum gravity in
three dimensions \cite{carlip}. In this sense, our kinematical formulation of the quantum theory
resembles maximally that of four dimensions, including potential nonperturbative ``entropic" effects coming from the
path integral measure, which have been shown to contribute to the effective quantum dynamics in 
four dimensions \cite{entropy}.

As we shall see below, our main aim -- to analyze the behaviour of shape observables in nonperturbative quantum 
gravity -- raises some interesting and qualitatively new issues on the way, which have to do with the role of 
boundary conditions and global topology. These are brought to the fore by a comparison of the dynamics of
{\it torus universes} with previous results in the
CDT formulation, which have almost all been derived for spherical spatial topology. 

This paper will analyze the quantum dynamics of volume in the new toroidal set-up, discussing in particular 
the issue of nontrivial boundary conditions.
The complementary description of the quantum dynamics of shape for the torus universes and their interaction with
the volume observable will be given in a companion paper \cite{toappear}. 

In the next section, we recall the most important elements of Causal Dynamical Triangulations in three spacetime
dimensions, and briefly review existing work on the subject. Sec.\ \ref{sec:torusuniverses} specializes to the case
of toroidal spatial topology and introduces the issue of boundary conditions and their relation to the volume profile. 
In Sec.\ \ref{sec:torusclassicalsolutions} we set the stage for a comparison with classical gravity, in the form 
of a minisuperspace
approximation depending only on the global scale factor and two global shape variables (or ``moduli"). 
Sec.\ \ref{sec:volumeprofiles} discusses the measurements of volume profiles, extracted from Monte Carlo 
simulations, and 
examines how they compare with the predictions of a generalized set of classical equations of motion.
In Sec.\ \ref{sec:volumecorrelations} we reconstruct
part of the effective action that governs the quantum dynamics of the scale factor, by measuring the
correlation matrix of volume fluctuations at different times. Finally, a summary and conclusions are presented in
Sec.\ \ref{sec:conclusions}.

\section{Causal dynamical triangulations in 2+1 dimensions}\label{sec:introductioncdttorus} 

Quantizing gravity in 2+1 dimensions is a popular toy model for the full, physical theory in 3+1 dimensions. 
The formulation of both theories in terms of the spacetime metric $g_{\mu\nu}(x)$ looks almost identical classically, but with the drastic 
simplification that there are no propagating {\it physical} degrees of freedom in $d= $2+1. Unlike what is usually done in
so-called ``reduced phase space quantizations" \cite{carlip}, we will make only limited use of the simplifying properties of gravity in
three dimensions when setting up the quantization. The concrete quantum-gravitational framework we will use, 
that of CDT, has the advantage of 
coming with a set of computational tools that allows us to access the nonperturbative sector of the model and evaluate
interesting observables numerically, with the help of Monte Carlo simulations. As mentioned earlier, our primary goal is
to learn about observables in nonperturbative quantum gravity, but we expect our investigation to lead at the same time to a 
better understanding of some specific features of the CDT approach. 

The logic we will be following is to define the quantum theory through a continuum limit of a regularized version
of the gravitational path integral, without putting in any preferred ``background geometry" and without assuming that
quantum fluctuations of geometry are necessarily small. Since in the present work both our simulations and the 
classical structures we will use for comparison lie in the Euclidean sector of the theory, we
will conduct the entire discussion in Euclidean (=Riemannian) metric signature (+++). While doing this, one should 
keep in mind that the
ensemble of geometries summed over in the Euclideanized CDT path integral is motivated by considerations
inherent to the Lorentzian, causal version of the theory before the Wick rotation, as has been discussed at length 
elsewhere \cite{cdtrev,physrep}.

Starting point is the formal continuum path integral on a three-dimensional manifold of product topology 
$[0,1]\times\Sigma$,
with initial and final spatial geometries on $\Sigma$ labelled by $g_{ab}$ and $g'_{ab}$,
\begin{equation}\label{eq:pathintegral}
Z[g_{ab},g'_{ab}]=\int \frac{\mathcal{D}g}{\mathrm{Diff}} \exp(-S_{EH}[g]),
\end{equation}
where the action is the Euclidean Einstein-Hilbert action
\begin{equation}\label{eq:ehaction}
S_{EH}[g_{\mu\nu}] = -\kappa\int \rmd^3x\sqrt{g}(R - 2\Lambda).
\end{equation}
The integral in (\ref{eq:pathintegral}) is over three-dimensional geometries, redundantly parametrized by metrics 
$g_{\mu\nu}$, which reduce to $g_{ab}$ and $g'_{ab}$ when restricted to the initial and final boundaries. 
Our notation is meant to indicate that the action of the diffeomorphism group $\mathrm{Diff}$, leading to this redundancy, 
must be factored out to remove infinite contributions and make sure that only physical configurations are counted.

Causal dynamical triangulations (CDT) is a particular regularization of this path integral which turns the 
infinite-dimensional integral into a discrete sum.
This is achieved by restricting the statistical ensemble underlying (\ref{eq:pathintegral}) 
to piecewise linear geometries of a specific form, where each geometry consists
of a ``stack" of $T$ thick slices. Each slice corresponds to a minimal time step and is assembled from 
three-dimensional simplices, solid tetrahedra whose interior geometry is flat.
Every path integral configuration therefore comes with a natural discrete time coordinate $t\in\{0,1,2,\dots ,T\}$, 
counting the number of thick slices. By construction,
the spatial geometries at fixed integer $t$ are of the form of two-dimensional triangulations $\mathbf{T}_t$ built from flat
equilateral triangles. This includes the two boundary geometries $g_{ab}$ and $g_{ab}'$, which are represented by
$\mathbf{T}_0$ and $\mathbf{T}_T$.

\begin{figure}[t]
\begin{center}
\includegraphics[height=4cm]{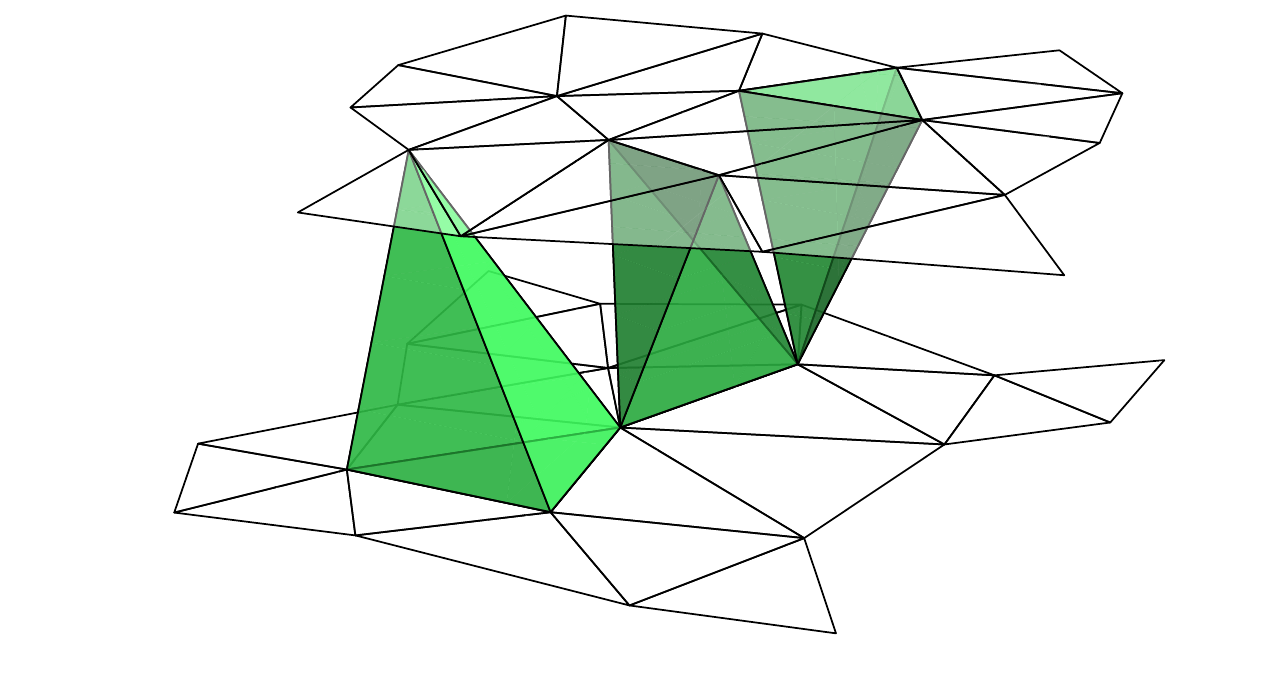}
\end{center}
\caption{The three different simplex types used in CDT in 2+1 dimensions: a 31-, a 22- and a 13-simplex,
labelled according to the numbers of vertices they share with the lower and upper spatial slice of integer $t$.
\label{fig:simplices}}
\end{figure}

The spacetime between each pair of adjacent spatial triangulations $\mathbf{T}_t$ and $\mathbf{T}_{t+1}$ 
is filled in with tetrahedra, in such a way as to make the three-dimensional geometry into a simplicial manifold.
In the standard CDT formulation, we allow three types of tetrahedra, called 31-, 22- and 13-simplices, according 
to the distribution of their vertices on consecutive spatial slices (see Fig.\ \ref{fig:simplices}).
All tetrahedra of a particular type are geometrically identical, which is implemented in CDT by making 
two choices, one for the length of all timelike edges (those connecting consecutive integer-$t$ slices), and
one for the length of all spacelike edges, those contained entirely in the spatial triangulations. 

\begin{figure}[t]
\begin{center}
\includegraphics[height=4cm]{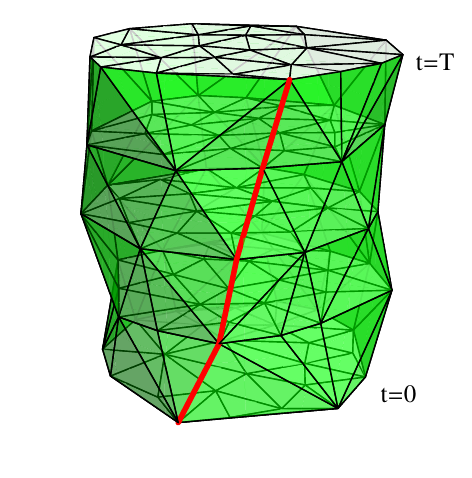}
\end{center}
\caption{If one uses ``edge distance" as a distance measure on causal dynamical triangulations, any vertex on the final 
boundary of a CDT geometry has distance $T$ to the initial boundary. \label{fig:constantdistance}}
\end{figure}

One consequence of using identical building blocks is that the geometry can be described essentially in combinatorial terms:
to specify a triangulation one only needs to keep track of a finite list of numbers describing the neighbourhood relations 
among the simplices, for example, in the form of an adjacency matrix. 
This way the ensemble of geometries in the path integral (\ref{eq:pathintegral}) becomes a discrete set 
$\mathcal{T}$ of three-dimensional triangulations $\mathbf{T}$.
The corresponding partition function for CDT is given by
\begin{equation}\label{eq:cdtpartitionfunction}
Z_{\mathrm{CDT}}[\mathbf{T}_0,\mathbf{T}_T,T] = \sum_{\mathbf{T}\in\mathcal{T}} \frac{1}{C_{\mathbf{T}}} e^{-S_{\mathrm{CDT}}[\mathbf{T}]},
\end{equation}
where $C_{\mathbf{T}}$ is the order of the automorphism group of the triangulation $\mathbf{T}$ and $S_{\mathrm{CDT}}$ is the 
Euclidean Einstein-Hilbert action evaluated on a piecewise linear manifold, now expressed as a function of the 
combinatorial properties of $\mathbf{T}$. One finds \cite{npb2001,3dcdt} that the action depends on 
the numbers $N_0$ of vertices and $N_3$ of tetrahedra contained in $\mathbf{T}$ according to 
\begin{equation}\label{eq:cdtaction}
S_{\mathrm{CDT}}[\mathbf{T}] = k_3 N_3[\mathbf{T}] - k_0 N_0[\mathbf{T}].
\end{equation}
The couplings $k_0$ and $k_3$ can be expressed in terms of the bare Newton and cosmological constants as well as
the space- and timelike edge lengths, but their precise form is of little interest here.
The important point is that the Einstein-Hilbert action yields a function linear in the numbers $N_d$ of simplices of 
dimension $d\leq 3$. $S_{\mathrm{CDT}}$ is the most general such expression, because of identities expressing 
the numbers $N_2$ of triangles and $N_1$ of edges 
in terms of $N_0$ and $N_3$. The same is true for the numbers of 31-, 22- and 13-simplices.
It also implies that the choice of spacelike and timelike edge lengths does not affect the CDT partition function other than 
by rescaling the bare Newton and/or cosmological constant.
One can view the preferred time foliation in CDT as a discrete analogue of a proper-time 
(or rather proper-distance) foliation of a Riemannian manifold.
Defining the \emph{edge distance} between two vertices as the minimal number of edges connecting them, 
any vertex in the spatial triangulation $\mathbf{T}_t$ has a fixed edge distance $t$ to the initial boundary.
In particular, both boundaries are separated by a fixed distance $T$ in lattice units, as illustrated by Fig.\ \ref{fig:constantdistance}.
The discrete parameter $t$ provides a convenient label that can be used in the construction of some observables, 
like the volume-volume correlators considered in Sec.\ \ref{sec:volumecorrelations} below.
Whether or not it assumes the role of a physically distinguished notion of time in the continuum limit of the theory is
not clear a priori.\footnote{A generalization of CDT quantum gravity, which does not have a distinguished time-slicing, has
recently been investigated in 2+1 dimensions \cite{cdtgeneral}.} 
The above-mentioned matching of CDT volume profiles to continuum de Sitter universes certainly 
suggests that an appropriately rescaled version of $t$ can be identified with proper time on large scales and ``on average" in 
this limit. Below in Sec.\ \ref{sec:volumeprofiles} we will try to match some of our results to a specific classical continuum
description with a distinguished proper time. 

The partition function $Z_{\mathrm{CDT}}$ can be used to define the expectation value of an observable 
$\mathcal{O}:\mathcal{T}\to\R$ according to
\begin{equation}
\langle \mathcal{O} \rangle = \frac{1}{Z_{\mathrm{CDT}}} \sum_{\mathbf{T}} \frac{\mathcal{O}
(\mathbf{T})}{C_{\mathbf{T}}} e^{-k_3 N_3 + k_0 N_0}.
\end{equation}
In Monte Carlo simulations of CDT we can measure these expectation values for certain observables.
Note that the transition amplitudes $Z[\mathbf{T}_0,\mathbf{T}_T]$ {\it as function of the 
boundary geometries} are in general {\it not} the most straightforward objects to study and interpret. This has to do with our
incomplete knowledge of the Hilbert space underlying the continuum theory, 
and how states labelled by discrete data $\mathbf{T}$ relate to continuum wave functions ``$\Psi(g_{ab})$" depending
on metric data. In a nonperturbative setting like the one we are considering, ``typical" states will have little resemblance
with semiclassical objects, certainly not on short scales, in the same way as ``typical" path integral histories do not
resemble classical spacetimes. Using specific boundary geometries $\mathbf{T}_0$ and $\mathbf{T}_T$ constructed
by hand runs the risk of introducing a bias in the simulations which we currently have no control over. 
One standard way of dealing with this issue is to avoid boundaries altogether by making time periodic, such that the 
topology of the triangulation becomes $S^1 \times \Sigma$. An alternative we will be using below is to work with a set
of singular boundary conditions of ``big bang" or ``big crunch" type, where the spatial volume shrinks to zero. This
leads to a drastic reduction in the number of variables needed to describe $\mathbf{T}_0$ and $\mathbf{T}_T$ and 
therefore to a situation which is much better controlled.

\begin{figure}[t]
\begin{center}
\includegraphics[height=5cm]{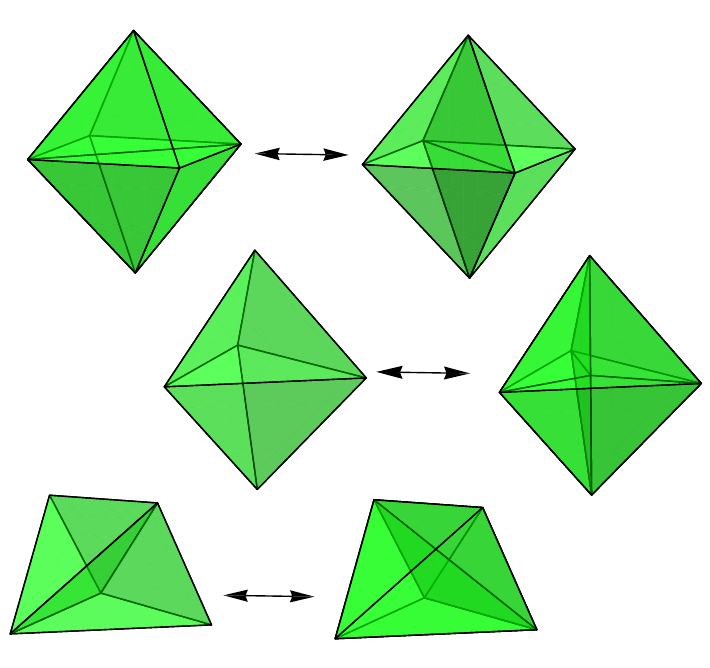}
\end{center}
\caption{A set of local update moves on CDT configurations in 2+1 dimensions:
flipping the diagonal of the central spatial square (top); 
subdividing the central spatial triangle into three (middle);
substituting the central timelike triangle by its dual timelike link (bottom).
\label{fig:moves}}
\end{figure}

To approximate expectation values of observables using Monte Carlo simulations one needs to generate a large set of 
random CDT configurations according to the Boltzmann distribution in (\ref{eq:cdtpartitionfunction}), which
can be accomplished by a Markov process. In practice,  we start by constructing by hand a triangulation $\mathbf{T}$ 
with the desired topology and satisfying the desired boundary conditions.
We then apply a large number of random update moves on $\mathbf{T}$, where each move occurs with a 
probability carefully chosen to satisfy a detailed balance condition.
In the case of CDT in 2+1 dimensions a suitable set of local update moves is shown in Fig.\ \ref{fig:moves}, 
see \cite{npb2001,3dcdt} for more details. We should point out that the Monte Carlo code used to derive the
results presented below is independent of what was used in previous published work, 
for example, in \cite{3dcdt,Benedetti:2009ge,kommu}. 

As already mentioned, properties of CDT quantum gravity in three spacetime dimensions have so far
been studied only for spherical spatial slices, that is, $\Sigma=S^2$. The phase structure of the underlying
statistical model was investigated in \cite{3dcdt,Ambjorn:2002nu}. The phase diagram is parametrized by the bare constants
$k_0$ and $k_3$ introduced above. As usual in dynamically triangulated systems, the bare cosmological constant 
(which in our case can be identified with $k_3$) must be fined-tuned from above to a critical line in phase space
to obtain an infinite-volume limit (divergent $N_3$).\footnote{Equivalently, the actual simulations 
are usually run at fixed system size $N_3$, and results for infinite volume are extracted by systematically studying
the scaling properties of the system as $N_3$ increases.} What remains is a one-dimensional
phase space parametrized by $k_0$. For a range of $k_0$-values, the system is found to be in a phase of 
extended, three-dimensional geometry, with a volume profile
that can be matched to that of a round three-sphere, the Euclidean de Sitter universe. As $k_0$ increases, one finds
a (first-order) phase transition to a phase of degenerate geometry, where neighbouring spatial slices decouple, and
which seems to be uninteresting from a physical point of view. An interesting feature of this system 
is that the quantum spacetime appears to be dynamically driven towards an $S^3$-topology,
although all geometries in the ensemble had product topology $S^1\times S^2$ (with time compactified to a circle) to start
with.

Studying a diffusion process on the CDT ensemble \cite{Benedetti:2009ge}, further evidence was gathered that 
the emergent quantum geometry on large scales is indeed a de Sitter universe. A measurement of the spectral 
dimension on short scales found a value compatible with 2 \cite{Benedetti:2009ge}, reminiscent of the 
dynamical dimensional reduction seen in CDT quantum gravity in 3+1 dimensions \cite{spectral}. Deviations from
sphericity of the {\it spatial} slices in the de Sitter phase of CDT have been studied in \cite{sachs}, using an
embedding in three-dimensional flat space to set up a spherical harmonic analysis. An generalized version
of CDT quantum gravity in 2+1 dimensions, including additional terms in the action motivated by Ho\v rava-Lifshitz
gravity, was investigated numerically in \cite{cdthorava}.

In terms of solving 2+1 Causal Dynamical Triangulations exactly, perhaps unsurprisingly for a three-dimensional 
statistical model, our knowledge is still rather incomplete. Matrix model methods
have been invoked to study the combinatorics of a single thick slice ($\Delta t=1$) and extract information
about the system's phase structure, transfer matrix and behaviour under renormalization \cite{mamo}.
One possible strategy for solving the gravity theory analytically is to reduce the number of CDT configurations 
to simplify the solution of the combinatorial problem without (hopefully) changing the universality class of the
continuum theory. By imposing additional order on spatial slices (more specifically, by requiring them to have
a product structure as two-dimensional triangulations) and by using an arsenal of statistical mechanical tools,
a Hamiltonian for the scale factor was derived analytically, for the case of cylindrical spatial slices \cite{blz}.
Using an even stronger restriction of the CDT ensemble by requiring spatial slices at integer-$t$ to be
flat tori, it could be shown that the combinatorics of the one-step propagator is that of a set of vicious walkers \cite{dl}. 
Despite the finite dimensionality of this model of ``CDT quantum cosmology", its continuum limit and full dynamics
are at this stage not known. It is a precursor of our present numerical investigation in the sense of using toroidal universes
and trying to extract a quantum dynamics in terms of three global parameters: the spatial volume and two moduli
parameters, describing the tori's global shape. -- Three-dimensional CDT quantum gravity with toroidal slices is
the subject we will turn to next.

\section{Torus universes in CDT}\label{sec:torusuniverses}

In this section we present an initial investigation of the spatial volume profiles in CDT simulations 
with $\Sigma=T^2$, and determine which boundary conditions yield the most interesting dynamics.
We will first consider periodic boundary conditions, which have been used extensively in CDT with spherical spatial 
topology. Note that periodic boundary conditions induce a time translation symmetry in the system:
shifting the time $t\to t+1$ (modulo $T$) maps the ensemble $\mathcal{T}$ of CDT configurations to itself and 
leaves the action $S_{\mathrm{CDT}}$ invariant. 
Consequently, the ensemble average $\langle V(t) \rangle$ of the spatial volume profile is time-independent and 
therefore contains little information.

\begin{figure}[t]
\begin{center}
\includegraphics[height=4.5cm]{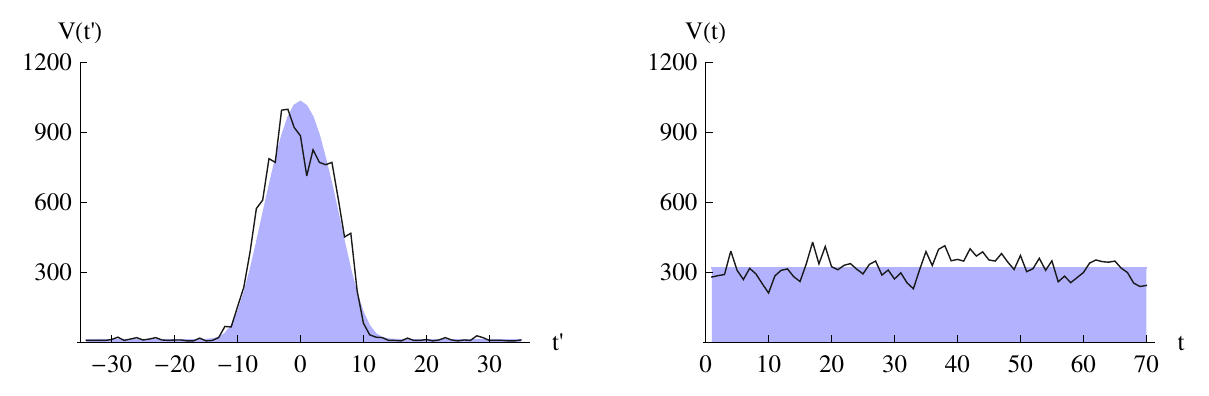}
\end{center}
\caption{Comparing volume profiles of CDT with topology $S^1\times S^2$ (left) and $S^1\times T^2$ (right). The solid curves 
represent typical (random) configurations, while the shaded areas correspond to the expectation values $\langle V(t')\rangle$ 
and $\langle V(t) \rangle$ respectively. (Sphere data taken at total volume $N_3\! =\! 40\, 000$, torus data at 
$N_3\! =\! 60\, 000$.)
\label{fig:periodicprofile}}
\end{figure}

However, it has been observed in the spherical case, both in 2+1 \cite{3dcdt} and 3+1 
dimensions \cite{ajl-prl}, that for individual spacetime configurations and sufficiently large time extent $T$ 
the simplices do not distribute themselves homogeneously in time, but ``condense" into a subinterval in time where
spatial volumes are macroscopic, while the remaining time slices are reduced to minimal spatial volume.
One can therefore obtain a nontrivial profile by averaging the spatial volumes not at fixed time $t$, but at a fixed time $t'$ 
relative to the location of the centre of the extended region along the time axis.
The resulting volume profile is illustrated in Fig.\ \ref{fig:periodicprofile} (left), which shows a typical volume distribution 
$V(t')$ together with the expectation value $\langle V(t')\rangle$.
To high accuracy the expectation value $\langle V(t')\rangle$ coincides with the spatial volume of a proper-time 
foliation of the three-sphere, lending credence to the conjecture that Euclidean de Sitter space emerges dynamically
from CDT quantum gravity on $S^1\times S^2$.

One might therefore have expected a similar behaviour for CDT on the torus, but according to our current understanding the
situation appears to be different. 
In none of the simulations performed, with a wide range of three-volumes $N_3\!\leq 250\, 000$, time extents 
$T\!\leq \! 100$, and couplings $k_0\!\in\! [2.5,4.5]$, did we observe a breaking of the time translation 
symmetry or any tendency of spatial slices to degenerate to configurations of minimal volume.
This is illustrated by the right-hand side of Fig.\ \ref{fig:periodicprofile}, which shows a typical volume profile from a 
CDT simulation with spatial tori and $T=70$ time slices. Of course, the absence of degenerate spatial geometries does 
not prove that the classical limit has time translation symmetry. A more detailed analysis, for instance, involving the 
distribution of Fourier modes of $V(t)$, would be necessary to make more definite statements.

At this stage, we do not know which feature of the torus topology in CDT is responsible for this quite different behaviour,
compared to that for spherical topology.
One way in which the CDT configurations differ is in the number of triangles in a minimal triangulation of $\Sigma$ (minimal 
in the sense of being compatible
with the simplicial manifold character and the overall topology of spacetime, i.e. preventing the geometry from ``pinching off").
For the sphere the minimal configuration is given by the boundary of a tetrahedron, which consists of four triangles.
By contrast, a triangulation of the torus requires a minimum of 14 triangles and therefore substantially 
more tetrahedra are needed to produce a ``stalk'' of minimal spatial volume. This could certainly have an effect on the simulations,
because the extra cost of the stalk (as measured by the $N_3$-term in the CDT action) may no longer be outweighed 
by the entropic gain of the tetrahedra clumping together. However, this is an effect one would expect to disappear for
sufficiently large three-volume. Once we have understood better the effective dynamics of CDT -- which is precisely
part of the motivation behind the present investigation -- we should be able to give a more satisfactory explanation of the 
global behaviour of the scale factor and its relation to topology.

The drawback of the apparent absence of symmetry breaking in time of the volume profiles is that it deprives us of
an interesting observable.
To bring about a nontrivial time dependence of the spatial volume,
we will in the following trade the usual periodic boundary conditions for suitable fixed boundaries, involving degenerate
boundary geometries of zero volume. For spherical topology such boundary conditions at $t=0$ and $t=T$ are not likely 
to have much of an effect, since the system wants to pinch off dynamically at the north and south pole of the de Sitter
universe anyway. By contrast, we will see that similar boundary conditions for CDT on the torus will have an impact 
on the dynamics of the spatial volume, since minimal spatial volumes do not occur with periodic boundary conditions.

\begin{figure}[t]
\begin{center}
\includegraphics[height=5cm]{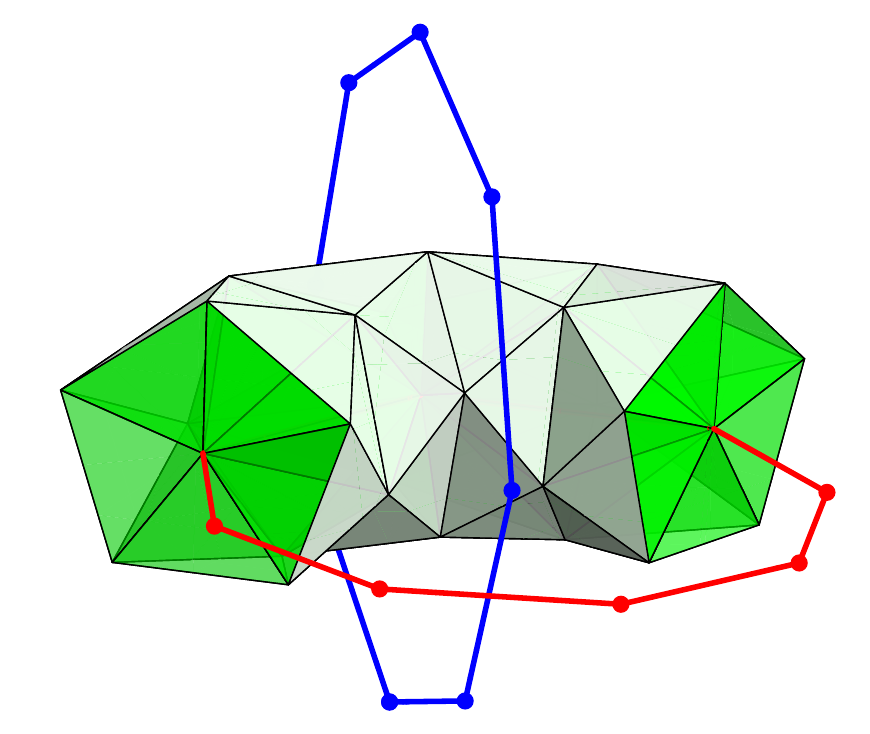}
\end{center}
\caption{A degenerate initial boundary at $t\! =\! 0$ consisting of $l_0\! =\! 8$ edges (solid red curve). Moving 
away from $t\! =\! 0$ in discrete time steps, one obtains a sequence of solid tori. The light triangles belong to the
boundary of the triangulated solid torus at $t\! =\! 1$. If we choose the final singularity according to the blue curve, 
we end up with the Hopf foliation of $S^3$.\label{fig:singularity}}
\end{figure}

In addition to a nontrivial volume profile we are also interested in creating a situation where the shape 
parameters of the tori evolve nontrivially in time. These two requirements can be satisfied simultaneously by taking an 
initial torus elongated in one direction and a final torus elongated in the opposite direction.
The difficulty with choosing boundary triangulations of this kind is that (i) we need many boundary triangles, and 
(ii) there is still considerable ambiguity in how we put them together.
Fortunately there is an easier option, namely, to take the boundary to be completely degenerate in one of the two directions, 
resulting in the collapse of the two-dimensional torus to a one-dimensional circle, described merely in terms of the 
number $l_0$ of its edges. To illustrate the idea, Fig.\ \ref{fig:singularity} shows a piece of a CDT configuration with 
an initial boundary consisting of $l_0=8$ edges.  

With this choice, the topology of the spacetime region $0 \leq t \leq t'$ for some $0<t'<T$ becomes that of a solid torus. 
If we impose similar boundary conditions on the final boundary, the same will hold for the region $t'\leq t \leq T$.
Therefore the most general spacetime topology is that of a pair of solid tori glued along their boundaries 
(in this case corresponding to the torus at time $t'$). 
This can be done in several topologically inequivalent ways, giving rise to $S^2\times S^1$, $S^3$ or, more generally, 
a \emph{lens space} $L(p,q)$ (see, for example, \cite{hatcher}).
For our purposes the second option is the most interesting, because it is the simplest topology that allows for 
a nontrivial shape evolution. It is achieved by taking the initial and final singularity such that together they form 
the so-called \emph{Hopf link} in $S^3$.
This is illustrated in Fig.\ \ref{fig:singularity}, where the final singularity is shown in blue and the embedding space 
$\R^3$ represents $S^3$ with one point removed (for example, after stereographic projection).
The foliation of the three-sphere by tori obtained in this way is known as the \emph{Hopf foliation}.

\begin{figure}[t]
\begin{center}
\includegraphics[height=5cm]{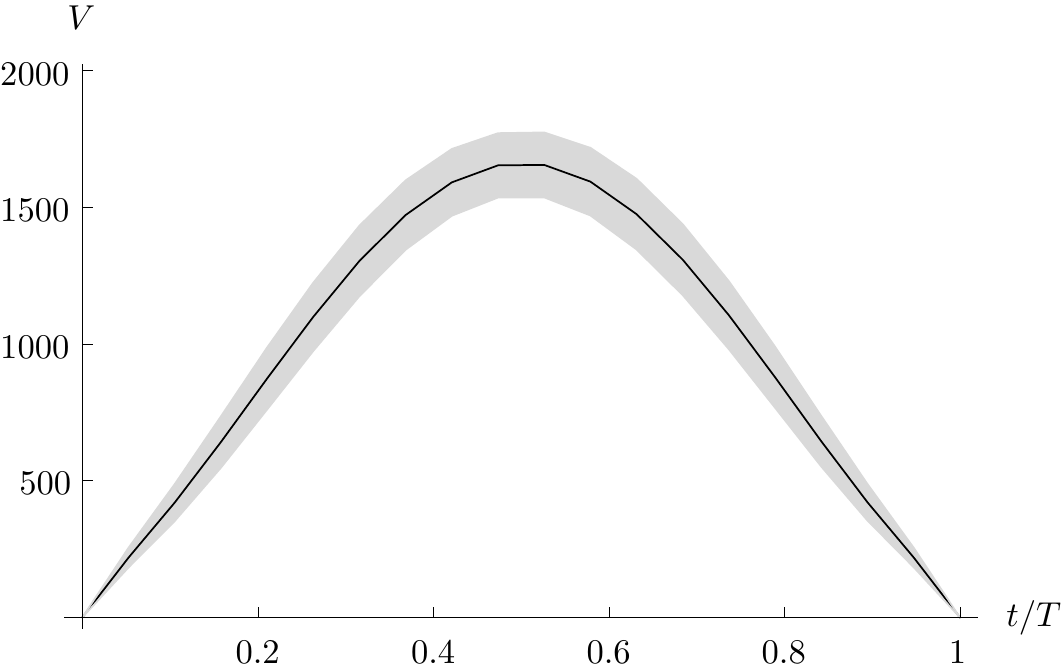}
\end{center}
\caption{Volume profile for $N_3\! =\! 60\,000$ and $T\! =\! 19$, with boundaries consisting of 
$l_0\! =\! l_1 \! =\! 60$ edges and $k_0\! =\! 2.5$. 
The shaded area corresponds to the standard deviation in $V(t)$ and gives an idea of the size of the quantum 
fluctuations. Error bars are not shown explicitly, but are of the order of $0.1\%$. \label{fig:volumefluc}}
\end{figure}

The results presented below are based on CDT simulations with time extension $T=19$. Unless indicated 
otherwise, we take the length $l_1$ of the final singularity at $t=T$ to be identical to the length $l_0$ of the 
initial singularity at $t=0$, in order to maintain time-reversal symmetry.
Fig.\ \ref{fig:volumefluc} shows the expectation value $\langle V(t)\rangle$ of the spatial volume, for a simulation 
with a fixed number $N_3=60\,000$ of tetrahedra, boundary length $l_0=60$, and coupling $k_0=2.5$.
We observe a clear expansion of the volume at early times and a contraction at late times, indicating 
a nontrivial dynamics of the spatial volume. Moreover, the expansion close to the singularity is roughly linear, 
which is in accordance with the initial geometry being one-dimensional.  
Before exploring other values of $N_3$, $l_0$, and $k_0$ more systematically, 
let us survey which classical gravitational solutions we may want to compare our results to.

\section{Classical solutions with torus topology}\label{sec:torusclassicalsolutions}

Classical solutions of General Relativity with Euclidean signature are given by the stationary points 
of the Euclidean Einstein-Hilbert action (\ref{eq:ehaction}), determined by solutions to the Einstein equations
\begin{equation}
R_{\mu\nu}-\frac{1}{2}R\, g_{\mu\nu}+\Lambda\, g_{\mu\nu} = 0,
\end{equation}
which in three dimensions are equivalent to (see, for example, \cite{carlip})
\begin{equation}
R_{\mu\nu\rho\sigma} = \Lambda(g_{\mu\rho}g_{\nu\sigma} - g_{\mu\sigma}g_{\nu\rho}).
\end{equation}
In other words, the solutions have constant scalar curvature $R$ and are therefore locally isometric to the 
three-sphere ($\Lambda\! >\! 0$), flat Euclidean space ($\Lambda\! =\! 0$) or hyperbolic space ($\Lambda\! <\! 0$), 
depending on the sign of the cosmological constant $\Lambda$.
As a consequence, classical General Relativity in three dimensions has no local degrees of freedom.

To find its solutions explicitly, we switch to the ADM formalism and rewrite the metric $g_{\mu\nu}$ in terms 
of the spatial metric $g_{ab}$, the shift vector $N^a$ and the lapse function $N$ as
\begin{equation}
ds^2=N^2 dt^2+g_{ab}(dx^a +N^a dt)(dx^b+N^b dt).
\end{equation}
In terms of these the Einstein-Hilbert action reads
\begin{equation}\label{eq:admaction}
S_{ADM}[g_{ab},N^a,N] = -\kappa\int \rmd t\int \rmd^2x\sqrt{{}^{(2)}\! g}\, N\left(K^2-K_{ab}K^{ab} + {}^{(2)}\! R-2\Lambda\right),
\end{equation}
where ${}^{(2)}\! g$ is the determinant of the spatial metric, ${}^{(2)}\! R$ the two-dimensional scalar curvature and $K_{ab}$ 
the extrinsic curvature tensor
\begin{equation}
K_{ab} = \frac{1}{2N}\left( \dot{g}_{ab} - \nabla_a N_b -\nabla_b N_a\right).
\end{equation}
The only difference with the usual Lorentzian signature case is a plus instead of a minus sign in front of the 
potential term $({}^{(2)}\! R-2\Lambda)$ in the action, relative to the kinetic term $(K^2-K_{ab}K^{ab})$, which remains
unchanged.
If one puts the lapse $N$ to 1, the set of constant-$t$ surfaces defines a proper-time foliation of the spacetime manifold. 
To make the gauge choice $N\! =\! 1$ therefore seems particularly suggestive when comparing to CDT.

One can find the classical solutions by putting (\ref{eq:admaction}) into canonical form (see \cite{moncrief} 
or \cite{carlip}) and imposing the constant mean curvature (CMC) gauge condition in which one can solve the 
dynamics completely.\footnote{Of course, the situation is slightly different than usual: since we consider 
Euclidean instead of Lorentzian gravity, we cannot be sure to capture all possible solutions this way. 
This is not a real problem, since we are only interested in a limited class of solutions matching our boundary conditions.}
In this gauge the classical solutions for the torus can be shown to be spatially flat. 
The lapse only depends on time, while the shift can be chosen to vanish, which means that on shell the 
foliation fixed by the CMC gauge is a proper-time foliation, up to a rescaling of the time variable.

It follows that in general all solutions can be obtained from a minisuperspace model, where spatial homogeneity 
is imposed from the outset. To achieve this, let us set $N\! =\! N(t)$, $N^a\! =\! 0$ and $g_{ab}(t)\! =\! 
V(t)\hat{g}_{ab}(\tau_i(t))$, 
where $\hat{g}_{ab}(\tau_i)$ denotes the flat unit-volume metric on the torus parametrized by the two real 
moduli $\tau_1$ and $\tau_2$,
\begin{equation}\label{eq:bgmetricfromtau}
\hat{g}_{ab}(\tau) = \frac{1}{\tau_2} \begin{pmatrix} 1 & \tau_1 \\ \tau_1 & \tau_1^2+\tau_2^2 \end{pmatrix}.
\end{equation}
Substituting this ansatz into (\ref{eq:admaction}) we obtain the minisuperspace action
\begin{equation}\label{eq:minisuperspaceaction}
S[V,\tau_i,N] = \kappa \int \rmd t \left(\frac{1}{2N}\left(-\frac{\dot{V}^2}{V}+V\frac{\dot{\tau}_1^2+\dot{\tau}_2^2}
{\tau_2^2}\right) + 2 N \Lambda V\right).
\end{equation}
As desired, this is now a function of global scale (the two-volume $V$) and global shape (in the form of the $\tau_i$). 
Note that there is no contribution coming from the Ricci scalar ${}^{(2)}\! R$, because its integral over
the torus vanishes by virtue of the Gauss-Bonnet theorem.
To find the classical solutions, we identify two conserved quantities $\cal E$ and $p$, given by
\begin{align}
{\cal E} &= \frac{1}{2N}\left(-\frac{\dot{V}^2}{V}+V\frac{\dot{\tau}_1^2+\dot{\tau}_2^2}{\tau_2^2}\right)
-2 N \Lambda V, \label{eq:energydef}\\
p &= \frac{V}{N}\frac{\sqrt{\dot{\tau}_1^2+\dot{\tau}_2^2}}{\tau_2}.\label{eq:taumomentadef}
\end{align}
Moreover, variation with respect to the lapse $N(t)$ yields the initial value condition ${\cal E}\! =\! 0$. 
Imposing the proper-time gauge $N\! =\! 1$, one easily finds the most general solution for the spatial volume $V(t)$ 
up to time translation and time reversal,
\begin{equation}\label{eq:vsol}
V(t) = \begin{cases} \frac{p}{2\sqrt{\Lambda}}\sin(2\sqrt{\Lambda}\,t) & \text{if } \Lambda > 0\\ 
p\,t& \text{if } \Lambda = 0 \hskip2cm ({\cal E}\! =\! 0)\\
\frac{p}{2\sqrt{-\Lambda}}\sinh\left(2\sqrt{-\Lambda}\,t\right)& \text{if } \Lambda < 0.\end{cases}
\end{equation}
In addition, we have the static solution $\dot{V}\! =\! p\! =\! 0$ for $\Lambda\! =\! 0$\footnote{Note that General Relativity 
with spherical spatial topology does not have such a static solution due to the presence of a spatial curvature term, 
which could explain the breaking of time translation symmetry in this case. However, this relies on the initial value 
condition ${\cal E}\! =\! 0$, which we will discuss further below.} and  
exponentially expanding/contracting solutions $V(t)\! =\! \exp\left(\pm 2\sqrt{-\Lambda}t\right)$ and $p\! =\! 0$ for 
$\Lambda\! <\!  0$.

From (\ref{eq:vsol}), the only solution with both an initial and a final singularity, at $t\! =\! 0$ and 
$t\! =\! T\! =\! \pi/(2\sqrt{\Lambda})$ respectively, is obtained when $\Lambda\! >\! 0$. The corresponding 
general solution for the modular parameter $\tau\! =\! \tau_1 + i \tau_2$ traces out   
a geodesic in the Poincar\'e upper half-plane (the Teichm\"uller space of a genus-1 surface), whose speed 
is determined by $p$ in (\ref{eq:taumomentadef}).
Reaching a big bang or big crunch singularity of the spacetime corresponds to $\tau$ hitting the boundary of 
Teichm\"uller space, which is associated with degenerate tori.
The boundary conditions $\tau(0)\! =\! 0$ and $\tau(T)\! =\! i\infty$ give precisely the Hopf foliation of 
the three-sphere, described in Sec.\ \ref{sec:torusuniverses} above.
The general solution with these boundary conditions is parametrized by the lengths $l_0$ and $l_1$ of the singularities 
and is given by the spacetime metric
\begin{equation}\label{eq:solposlambda}
ds^2 = dt^2 + l_0^2 \cos^2(\sqrt{\Lambda}\,t)\, dx^2+ l_1^2 \sin^2(\sqrt{\Lambda}\,t)\, dy^2.
\end{equation}
The three-volume $V_3$ of this geometry is $V_3\! =\! l_0 l_1 /(2\sqrt{\Lambda})$, which directly relates the cosmological 
constant $\Lambda$ and the total volume $V_3$. One also finds $p\! =\! l_0 l_1 \sqrt{\Lambda}$.
However, when trying to compare the corresponding volume profile
\begin{equation}\label{eq:grvolumeprofile}
V(t)=\frac{l_0l_1}{2}\sin(2\sqrt{\Lambda}\,t)
\end{equation}
directly to measurements in CDT simulations, one runs into a difficulty.
The total time extension $T\! =\! \pi/(2\sqrt{\Lambda})$ of the classical solution (\ref{eq:grvolumeprofile}) 
is fixed in terms 
of $\Lambda$ (equivalently, in terms of the three-volume and the boundary conditions), whereas
in the CDT simulations $T$ appears a priori as an additional, free parameter set by hand.

This is a specific case of a more general issue, namely, under what circumstances
particular CDT set-ups and their associated observables can be meaningfully compared to aspects 
of the {\it classical} theory of General Relativity, and vice versa. To start with, there is of course no guarantee 
that an arbitrary classical solution can be obtained from a nonperturbative path integral\footnote{more precisely, 
since we are working in Euclidean signature,
as the minimum of some effective Euclidean action governing the quantum dynamics}, with suitable boundary conditions. 
Returning to the problem at hand, one way of trying to address the mismatch between the parameters of the 
classical torus solutions and those of the CDT model with specific boundary conditions imposed is to generalize the
classical solutions one is comparing to. The idea is to gauge-fix the lapse in the minisuperspace action 
(\ref{eq:minisuperspaceaction}) {\it before} varying the action, without adding the (then missing) Hamiltonian constraint 
by hand afterwards. As we will see, one gains a free parameter by doing this, and can write down a set of classical 
solutions which satisfy the desired boundary conditions. In the next section we will examine how well these
generalized solutions fit the measured simulation data. 

Whether or not such an ansatz captures the nonperturbative dynamics of the 2+1 CDT model with nontrivial boundaries
correctly is difficult to argue on general grounds, because of our lack of explicit control over the details of the former.
Apart from the already mentioned influence of boundary conditions and global topology, this includes the question
to what extent the discrete ``edge distance" $t$ on CDT configurations described in Sec.\ \ref{sec:introductioncdttorus}
in a particular continuum limit can be related to any continuum notion of a local (proper) time, which one can identify on a 
{\it classical} ensemble of spacetime metrics, through gauge-fixing or otherwise.
Although the de Sitter results are strong evidence that such an identification works on
large scales for the global volume variable, this does not imply that these two notions of time can be identified
at a local, microscopic level.

Keeping these comments in mind, we will now construct an ensemble of smooth classical metrics in proper-time
gauge, each of whose members has an initial and a final boundary at exact proper-time distance $T$, together with
a set of generalized equations for the global scale and shape variables. In line with our previous discussion, 
our starting point will be the Einstein-Hilbert action in ADM form, with the lapse fixed to $N\! =\! 1$, yielding
\begin{equation}\label{eq:actionfixedT}
S[g_{ab},N^a] = -\kappa \int_0^T \rmd t \int \rmd x \sqrt{g}\left(K^2-K_{ab}K^{ab} + R - 2\Lambda\right).
\end{equation}
This means that 
\begin{equation}\label{eq:hconstr}
\left.\frac{\delta S}{\delta N}\right|_{N=1} = -\sqrt{g}(K_{ab}K^{ab} - K^2 + R - 2\Lambda)
\end{equation}
is no longer required to vanish, although its constancy in time is still guaranteed by the other equations of motion.
Note that the functional form of the right-hand side of (\ref{eq:hconstr}) is exactly that of the Hamiltonian
constraint (in Euclidean signature), whose vanishing is normally used to solve 
for the trace part of the spatial metric in terms of the traceless degrees of freedom.

Let us determine which effect this has on the family of homogeneous solutions for the torus universe.\footnote{Contrary 
to the general relativistic case, homogeneity is a nontrivial restriction on the full set of solutions to (\ref{eq:actionfixedT}). 
There is an infinite-dimensional family of classical solutions due to the presence of a local degree of freedom. 
Only when the boundary conditions are homogeneous, which is the case we are interested in, can we safely 
assume homogeneity of the solutions.} We still have the conserved quantities $\cal E$ and $p$ from (\ref{eq:energydef}) 
and (\ref{eq:taumomentadef}), with $N$ set to 1. However, $\cal E$ is no longer required to vanish and serves as an 
additional parameter in the family of solutions, which we can tune to arrive at a desired total time extent $T$.
Restricting to the solutions with two singularities and non-vanishing $p$, we find a family of volume profiles,
\begin{equation}
V(t) = \begin{cases} l_0 l_1 \frac{\sinh\left(\sqrt{-\Lambda} (T-t)\right)\sinh\left(\sqrt{-\Lambda}\,t\right)}
{\sinh^2\left(\sqrt{-\Lambda}\,T\right)} & \text{if } \Lambda < 0 \\
l_0 l_1 (T-t)t/T^2 & \text{if } \Lambda = 0  \hskip2.5cm ({\cal E}\! \not= \! 0)\\
l_0 l_1 \frac{\sin\left(\sqrt{\Lambda} (T-t)\right)\sin\left(\sqrt{\Lambda}\,t\right)}
{\sin^2\left(\sqrt{\Lambda}\,T\right)} & \text{if } 0<\Lambda < \left(\frac{\pi}{T}\right)^2.\end{cases}
\label{eq:classicalvolumeprofiles}
\end{equation}
The underlying line element, say, for the case of positive $\Lambda$, generalizes (\ref{eq:solposlambda}) to
the four-parameter family
\begin{equation}\label{eq:solgenposlambda}
ds^2 = dt^2 + l_0^2\, \frac{\sin^2(\sqrt{\Lambda}\,(T-t))}{\sin^2(\sqrt{\Lambda}\, T)}\, dx^2+ 
l_1^2\, \frac{\sin^2(\sqrt{\Lambda}\, t )}{\sin^2(\sqrt{\Lambda}\, T)}\, dy^2.
\end{equation}
\begin{figure}[t]
\begin{center}
\includegraphics[height=5.5cm]{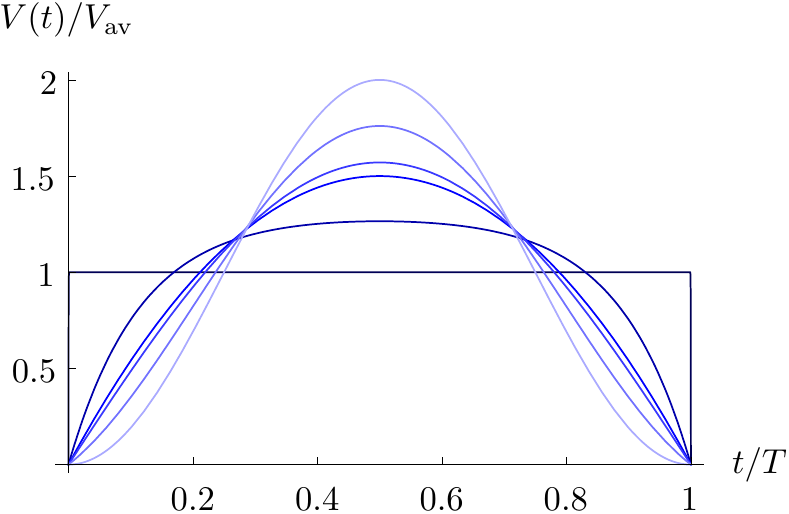}
\end{center}
\caption{The spatial volume $V(t)$ of (\ref{eq:classicalvolumeprofiles}) of the classical solutions, 
normalized by its time average 
$V_{\mathrm{av}}\! =\! V_3/T$, for different values of $v\! =\! V_3/(l_0l_1T)$. The value of $v$ increases 
monotonically from 0 to $\infty$ as $\Lambda$ ranges between $-\infty$ and $(\pi/T)^2$ (for fixed $l_0$,
$l_1$ and $T$). The curves, from dark to light, correspond to $v\! =\! 0$ (flat), $v\! =\! 0.01$, $v\! =\! 1/6$ (parabola, 
$\Lambda\! =\! 0$), $v\! =\! 1/\pi$ (sine, $\Lambda\! =\! (\pi/2T)^2$), $v\! =\! 2$, and $v\! =\! \infty$ (sine squared,
$\Lambda\! =\! (\pi/T)^2$).\label{fig:volumeplot}}
\end{figure}

\noindent The two associated constants of motion can be expressed as functions of the four parameters 
$(l_0,l_1,\Lambda, T)$ according to
\begin{equation}\label{eq:constgen}
{\cal E}=2\Lambda l_0 l_1\, \frac{\cos(\sqrt{\Lambda} T)}{\sin^2(\sqrt{\Lambda} T)}, \;\;\;
p= \frac{l_0 l_1 \sqrt{\Lambda}}{\sin (\sqrt{\Lambda} T)}. \;\;\;\;\;\;\;\;\;\; \left( 0<\Lambda < (\pi/T)^2\right)
\end{equation}
For fixed $l_0$, $l_1$, and $T$, the three-volume computed from (\ref{eq:classicalvolumeprofiles}) 
increases monotonically 
as a function of $\Lambda$ from $V_3\! =\! 0$ at $\Lambda\! =\! -\infty$ to $V_3\! =\! \infty$ at $\Lambda\! =\! (\pi/T)^2$.
In Fig.\ \ref{fig:volumeplot} we have plotted the analytic volume profiles normalized by their time average 
$V_{\mathrm{av}}\! =\! V_3/T$ for various values of $\Lambda$.
The shape of the profile only depends on the dimensionless quantity $v\! =\! V_3/(l_0 l_1 T)$. 
For $v\! \to\! 0$ we find a flat profile, for $v\! =\! 1/6$ a parabola, for $v\! =\! 1/\pi$ a sine, and for $v\! \to\!\infty$ a
sine-squared profile. Note also that the minisuperspace action (for $N\! =\! 1$),
when evaluated on solutions, is positive, despite the negative kinetic term for the volume in 
(\ref{eq:minisuperspaceaction}). Like the three-volume,
the action is monotonically increasing as a function of $\Lambda$ (with everything else held fixed), from
$S\! =\! 0$ at $\Lambda\! =\! -\infty$ to $S\! =\! +\infty$ for $\Lambda\! =\! (\pi/T)^2$.

\section{Measurement of volume profiles}\label{sec:volumeprofiles}

Having derived the volume profiles (\ref{eq:classicalvolumeprofiles}) for the generalized equations of motion,
we will now return to the data from the Monte Carlo simulations, to see whether they can be fitted to the new, wider
range of shapes. In the current set-up, the CDT partition function
depends on four free parameters (five if we counted the boundary lengths $l_0$ and $l_1$ separately): the time 
extension $T$, the discrete three-volume $N_3$, the coupling $k_0$, and the boundary length $l_0$.
Because probing the full parameter space would be very time-consuming, we fix the three-volume to 
$N_3=60\,000$ and the total time to $T=19$, and perform simulations for a wide range of values of the coupling 
$k_0$ and the boundary length $l_0$.

\begin{figure}[t]
\begin{center}
\includegraphics[height=5cm]{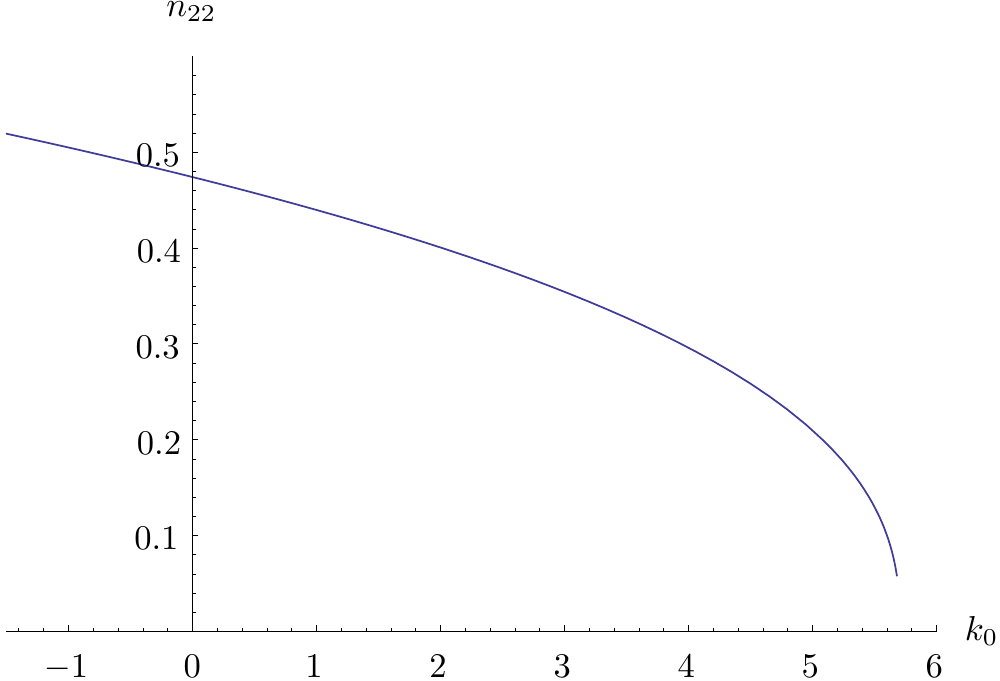}
\end{center}
\caption{The measured fraction $n_{22}$ of simplices of type $22$ as function of the 
coupling constant $k_0$. \label{fig:n22plot}}
\end{figure}

To determine the relevant range of the coupling $k_0$, recall from Sec.\ \ref{sec:introductioncdttorus} that 
$k_0$ -- which can essentially be identified with the bare inverse Newton constant -- is the free, tunable phase 
space parameter that remains after fine-tuning the bare cosmological constant. Of course, we expect the
phase diagram of 2+1 CDT quantum gravity on the torus to have the same qualitative features as on the sphere,
with an extended quantum spacetime only found below some value $k_0^{\ast}$. Like in the spherical
case \cite{3dcdt},
the effect of increasing $k_0$ is to reduce the number $N_{22}$ of 22-simplices in favour of 31- and 13-simplices.
As illustrated by Fig.\ \ref{fig:n22plot}, the fraction $n_{22}:=N_{22}/N_3$ of simplices of type 22 in our simulation 
collapses to (nearly) zero when $k_0$ approaches the critical value $k_0^{\ast}\approx 5.6$.
Since the 22-simplices provide the coupling between consecutive spatial triangulations, the 
spatial geometries in the region $k_0 > k_0^{\ast}$ of phase space effectively decouple, and spacetime 
loses any classical physical interpretation. Since we are interested in a theory which possesses a
good classical limit, and therefore is macroscopically extended in both time and space, 
we will from now on restrict our attention to the ``physical'' phase $k_0 < k_0^{\ast}$, and
also make sure to not get too close to the phase transition $k_0^{\ast}$, where the fluctuations in the spatial 
volume are large.

\begin{figure}[t]
\centering
\subfloat[][]{\includegraphics[height=5cm]{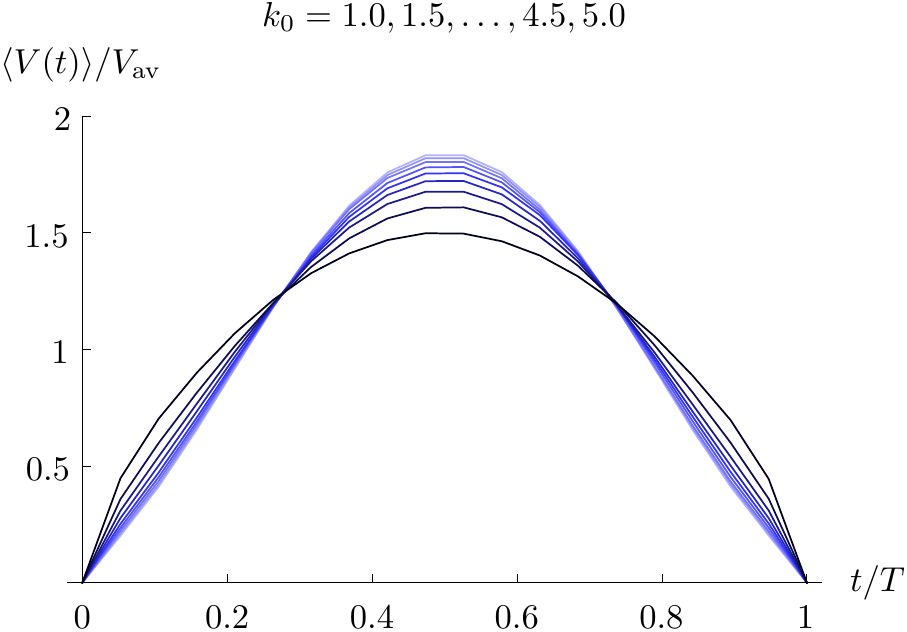}\label{fig:volumemeasurementk0}}%
\subfloat[][]{\includegraphics[height=5cm]{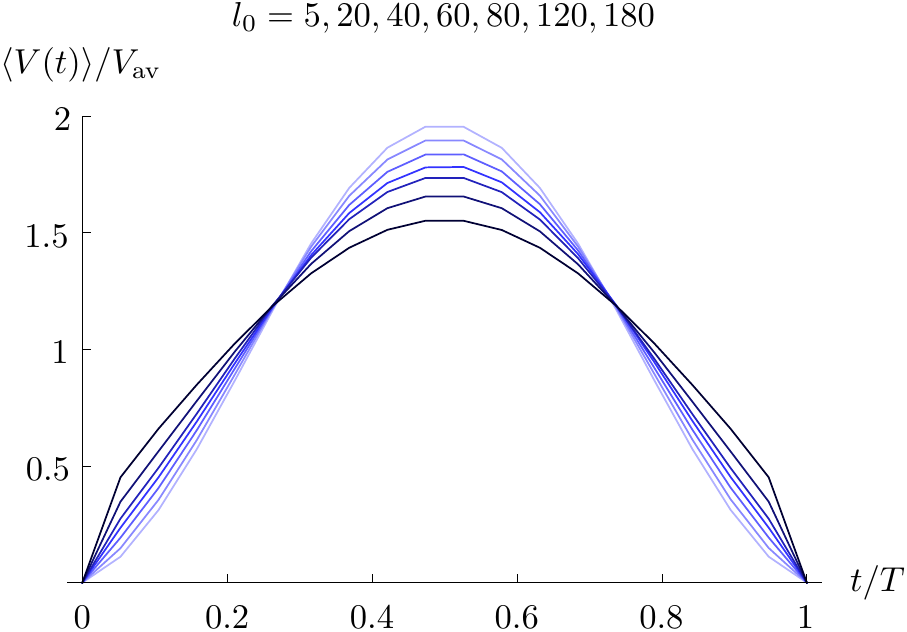}\label{fig:volumemeasurementl0}}
\caption{Normalized volume profiles for several simulations with $N_3\! =\! 60\,000$: 
(a) at fixed $l_0\! =\! 60$, for various values of $k_0$, and (b) at fixed $k_0\! =\! 2.5$, for various boundary 
lengths $l_0$. The lightest curves correspond to $k_0\! =\! 1.0$ and $l_0\! =\! 5$ respectively.}%
\label{fig:volumemeasurement}
\end{figure}

The expectation value $\langle V(t)\rangle$ presented earlier in Fig.\ \ref{fig:volumefluc} turns out to be 
well described by the volume profile for $v\! =\! 0.98$ (c.f. Fig.\ \ref{fig:volumeplot}). More systematically,
Fig.\ \ref{fig:volumemeasurementk0} shows our results for fixed $l_0\! =\! 60$ and $k_0$ varying from $1.0$ to 
$5.0$ in steps of $0.5$. Clearly, as we increase $k_0$ towards the phase transition, the shape of the volume profile 
becomes flatter, which corresponds to the parameter $v$ approaching zero.
This is in line with the discussion above, in the sense that a complete decoupling of the spatial triangulations 
would lead to a flat volume profile with $v\! =\! 0$.
Conversely, to obtain the curves in Fig.\ \ref{fig:volumemeasurementl0}, we fixed $k_0\! =\! 2.5$ but varied the 
boundary length $l_0$ to take the values 5, 20, 40, 60, 80, 120, and 180.
As $l_0$ decreases, we observe an approach towards the sine-squared shape, i.e. $v\to\infty$, in accordance with
the dependence of $v$ on the boundary lengths in the analytic continuum formulation.

If we were sufficiently confident that the system was well approximated by the classical minisuperspace description, 
we could at this stage test the classical relation $v\! =\! V_3/(T l_0^2)$ quantitatively.
However, we should probably not put too much trust in this classical description.
To start with, it is unclear how significant the qualitative similarity between the measurements and the classical solutions 
really is; it could be due to the rather generic nature of the classical volume profiles, which represent roughly the 
smoothest profiles with a given slope and time-reversal symmetry.
More significant tests of the conjectured classical limit would involve studying higher-order corrections to the volume 
profile.

\begin{figure}[t]
\begin{center}
\includegraphics[height=5cm]{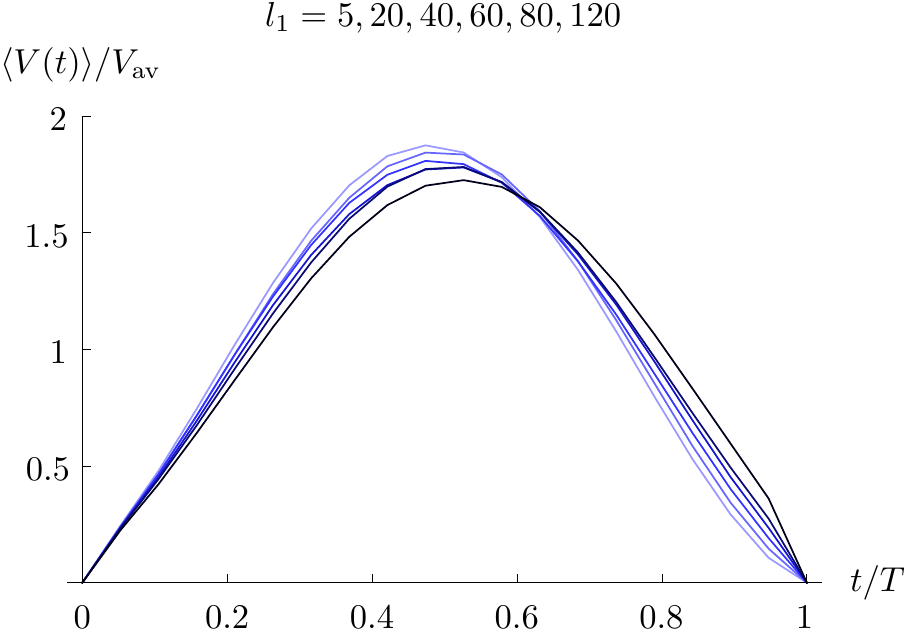}
\end{center}
\caption{Normalized volume profiles for $k_0\! =\! 2.5$, $N_3\! =\! 60\,000$ and $T\! =\! 19$. The initial singularity has 
fixed length $l_0\! =\! 60$, while the final singularity length is varied with $l_1\! =\! 5,20,40,60,80,120$ 
($l_1\! =\! 120$ corresponds to the darkest curve).\label{fig:volumemeasurementasym}}
\end{figure}

The situation is different when we consider CDT set-ups which are not symmetric under time reversal, 
by using unequal boundary lengths $l_0$ and $l_1$ in the simulations. In this case, 
the classical solutions (\ref{eq:classicalvolumeprofiles}) yield a nontrivial prediction, namely, that the volume 
profile remain symmetric. To test this prediction, we have performed simulations with fixed initial singularity length 
$l_0\! =\! 60$ and varying final singularity length $l_1\! =\! 5$, 20, 40, 60, 80 and 120.
The results are shown in Fig.\ \ref{fig:volumemeasurementasym}, from which it is clear that the expected symmetry 
is {\it not} present in our system when $l_0\! \neq\! l_1$.
Instead of identical slopes at the two boundaries, we see that the slope at the initial boundary hardly changes 
when changing $l_1$. If the classical solutions we are comparing to are the correct ones, 
this could mean that the information about the geometry at $t\! =\! T$ is not propagated all 
the way to small $t$, in the sense that spatial geometries at small time $t$ are oblivious to the final boundary conditions.
In classical gravity, however, the geometry is sensitive to the boundary conditions no matter how far one is from the boundary.

There are potentially a number of reasons which could explain discrepancies between the data and the classical solutions (\ref{eq:classicalvolumeprofiles}): (a) at least for certain boundary conditions, the classical limit of CDT could be 
genuinely different from the generalized minisuperspace ansatz we obtained from general relativity, (b) our systems 
could be too far from classicality or too small to make a sensible comparison, (c) we could be making the wrong 
identification of discrete and continuum boundary conditions; 
for example, the path determined by the singularity could exhibit a random walk behaviour in the three-dimensional 
triangulation, leading to an anomalous scaling of the continuum singularity length with the discrete $l_0$ (which
should be taken into account when investigating a dependence on boundary lengths, say).

At the system sizes we have been considering probably not much will be learned by making more 
detailed measurements of the volume profiles, since it is difficult to distinguish alternatives for the classical 
dynamics from quantum corrections.
As a potentially more fruitful strategy, we will from now on {\it assume} 
that for the range of system sizes we are considering, there exist effective actions which describe the CDT systems,
and will try to ``reconstruct" these effective actions or at least deduce some of their properties directly from the data.
Once we manage to significantly narrow down the relevant terms in a given effective action, we will in principle be 
able to study their scaling properties when approaching the continuum limit. In the next section, we will reconstruct
the kinetic term for the two-volume from measuring a particular correlation function. Forthcoming work will analyze
the contribution from the moduli parameters to the effective action \cite{toappear}.

\section{Effective action from volume correlations}\label{sec:volumecorrelations}

Suppose that the Euclidean effective action $S[V]$ for the spatial volume in 2+1 dimensional CDT is local 
in time and can be written as the integral of a Lagrangian $\mathcal{L}(V,\dot{V})$ according to
\begin{equation}\label{eq:genericeffactionv}
S[V] = \int_0^T \rmd t\,\mathcal{L}(V,\dot{V}).
\end{equation}
Assuming time reversal symmetry leads to the condition $\mathcal{L}(V,-\dot{V})=\mathcal{L}(V,\dot{V})$, 
which implies that only even powers of $\dot{V}$ can appear in the Lagrangian.
Given a proper set of boundary conditions, the action $S[V]$ will have a unique classical solution $V_0(t)$, 
satisfying $\delta S[V_0]\! =\! 0$, which should describe the expectation values $\langle V(t) \rangle$ of the
volume profile measured in the simulations, at least for sufficiently large $V(t)$ that are 
unaffected by quantum corrections.

Since in general $V_0(t)$ depends on all terms in the Lagrangian $\mathcal{L}$, it is difficult to deduce 
specific information about the form of $\mathcal{L}$ from measurements of $\langle V(t)\rangle$ alone.
It turns out that more specific information is contained in the quantum fluctuations around the classical solution.
Following a similar treatment in 3+1 CDT quantum gravity \cite{desitter}, assume that the fluctuations are small 
enough for a semiclassical treatment to make sense.
In that case the fluctuations $\delta V(t)\! =\! V(t) - \langle V(t)\rangle$ are correlated according to
\begin{equation}
\langle \delta V(t)\,\delta V(t')\rangle = \langle V(t)\,V(t')\rangle - \langle V(t)\rangle\langle V(t')\rangle 
\propto \left(\frac{\delta^2 S}{\delta V^2}[V_0]\right)^{-1}(t,t'),
\end{equation}
which means that one can deduce numerically the operator $P(t,t') = \frac{\delta^2 S}{\delta V^2}[V_0]$ 
by inverting the matrix of correlations of spatial volumes.

\begin{figure}[t]
\begin{center}
\includegraphics[height=5cm]{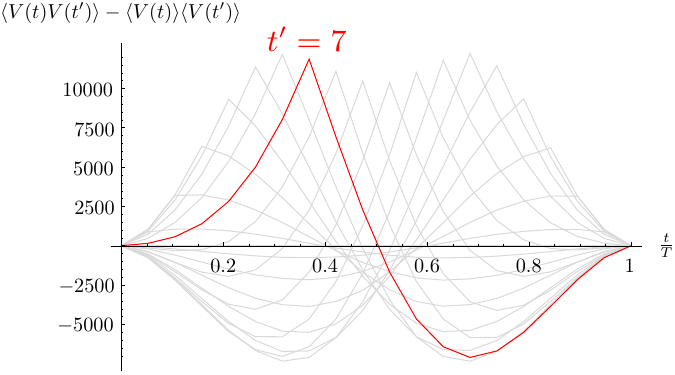}
\end{center}
\caption{Volume correlation function $\langle \delta V(t)\,\delta V(t')\rangle$ for a simulation with $T\! =\! 19$, 
$N_3\! =\! 70\,000$, $l_0\! =\! 75$, 
and $k_0\! =\! 1.2$. Individual curves correspond to various fixed values of $t'$; the red curve corresponds to 
$\langle V(t)V(7)\rangle\! -\!\langle V(t)\rangle\langle V(7)\rangle$. The occurrence of negative values is
due to the overall volume constraint. \label{fig:volumecor}}
\end{figure}

\begin{figure}[t]
\centering
\subfloat[][]{\includegraphics[height=4.5cm]{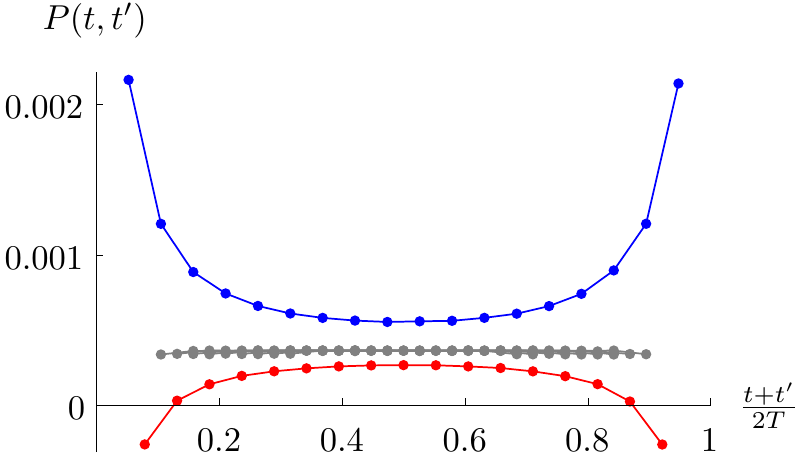}\label{fig:invcormat}}%
\subfloat[][]{\includegraphics[height=4.5cm]{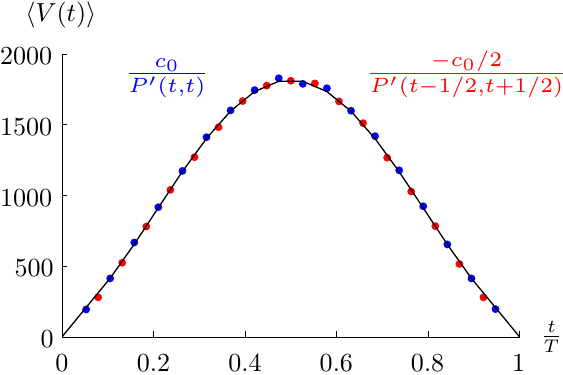}\label{fig:invcorcompare}}
\caption{(a) Elements of the inverse correlation matrix $P(t,t')$, linearly interpolated: on the diagonal 
(top, blue curve), on the first sub- and superdiagonal 
(bottom, red curve), and remaining off-diagonal elements (middle, grey; fat dots indicate accumulation of elements). 
(b) Volume expectation value $\langle V(t)\rangle$ (solid curve) compared to the diagonals of the normalized 
inverse correlation function $P'(t,t')\! =\! P(t,t')\! -\! P_0$. The fitted proportionality constant is $c_0\! \approx\! 0.35$.}%
\label{fig:invcorvolume}
\end{figure}

Fig.\ \ref{fig:volumecor} shows the measured volume correlations for a CDT simulation with 
$N_3\! =\! 70\,000$, $l_0\! =\! 75$, and $k_0\! =\! 1.2$.
Restricted to integer values $1\leq t,t' \leq T-1$, the correlation matrix $\langle \delta V(t)\,\delta V(t')\rangle$ is invertible.
When plotting elements of its matrix inverse $P(t,t')$, one observes that they fall naturally
into three groups, as indicated in Fig.\ \ref{fig:invcormat}: diagonal matrix elements lie on the
top curve, sub- and superdiagonal elements on the bottom curve, while all others matrix elements have approximately 
the same value, which we will denote by $P_0$.
This constant nonlocal contribution to the inverse correlations is due to the global constraint on the three-volume 
$N_3$, which enforces $\sum_t \delta V(t)\! =\! 0$. 
If one wanted to take this effect into account in the effective action (\ref{eq:genericeffactionv}), one should add a term 
of the form $f(\int_0^T \rmd t\,V(t))$. For convenience, we will simply subtract this constant term from 
$P(t,t')$ and work with the normalized inverse correlation matrix
\begin{equation}
P'(t,t') = P(t,t') - P_0.
\end{equation}

\noindent According to the ansatz (\ref{eq:genericeffactionv}), the continuum operator $P'$ is given by
\begin{equation}\label{eq:ansatzinvcor}
P'(t,t') = \left[ \frac{\partial^2\mathcal{L}}{\partial V^2} - 
2\, \frac{\rmd}{\rmd t} \left(\frac{\partial^2\mathcal{L}}{\partial V \partial \dot{V}}\right) - 
\frac{\rmd}{\rmd t} \left( \frac{\partial^2\mathcal{L}}{\partial \dot{V}^2}\left(\frac{\rmd}{\rmd t}\,\, \cdot\,\, \right)\right)\right]\delta(t-t'),
\end{equation}
where the partial derivatives of the Lagrangian are evaluated at $V\! =\! V_0(t)$.
The operator $P'$ consists of a purely diagonal part and a second-order time derivative,
whose time-dependent coefficient depends only on the kinetic term in the Lagrangian.

Comparing the numerical values of the elements on the first sub- and superdiagonal of the discrete normalized
operator $P'(t,t')$ with those on its diagonal (c.f. Fig.\ \ref{fig:invcormat}), they differ with high accuracy by a
factor of $-1/2$, as one would expect for the finite-difference representation of a second-order derivative.
We conclude that the matrix $P(t,t')$ represents a discretization of a second-order time derivative operator, 
like the last term in (\ref{eq:ansatzinvcor}). The purely diagonal component of the operator appears to be absent 
or small compared to the second-order time derivative part.
We can now try to extract the time-dependent prefactor $\partial^2 \mathcal{L}/\partial \dot{V}^2[V_0]$ of the 
kinetic term $\dot{V}^2$ in the effective action from the simulation data. It turns out that this prefactor is very 
close to $1/V_0(t)=1/\langle V(t)\rangle$, 
as shown in Fig.\ \ref{fig:invcorcompare}, where we have plotted the measured volume profile (solid curve) together 
with rescalings of the inverses of the diagonals from Fig.\ \ref{fig:invcormat} (red and blue dots).
The proportionality constant $c_0\approx 0.35$ has been obtained by a best fit.

We conclude that the volume-volume correlations are accurately described by a kinetic term in the effective action 
of the form
\begin{equation}\label{eq:effvolactionfromdata}
S[V(t)] = \int_0^T \rmd t \left( \frac{c_0}{2} \frac{\dot{V}^2}{V} + \cdots \right).
\end{equation}
This kinetic term is of the same form as the one appearing in the minisuperspace action \eqref{eq:minisuperspaceaction} 
(with $N=1$), up to a sign. The positive sign in (\ref{eq:effvolactionfromdata})
comes as no surprise, since the semiclassical treatment of the path integral 
relies on the fact that the classical solution $V_0(t)$ appears at a minimum of the Euclidean action.
(The minisuperspace action for the spatial volume has its classical solution at a saddle point.)
The reason for the sign difference of the kinetic term in the effective action extracted from the 
full CDT path integral can be traced back to nonperturbative ``entropic" contributions coming from the number 
of microscopic path-integral configurations (see, for example, \cite{entropy}).
Their appearance in dynamically triangulated formulations of quantum gravity is generic and they
play an important role in spacetime dimension $d\geq 3$ in rendering the effective action bounded from below 
and the Euclidean path integral stable\footnote{The Euclidean Einstein-Hilbert action is not bounded 
below, which leads to the well-known \emph{conformal mode problem}, at least in na\"ive implementations of 
the Euclidean path integral \cite{mazurmottola,dasguptaloll}.}, at least for suitably chosen bare coupling constants. 
In both three and four dimensions, the de Sitter universes found in the CDT approach emerge of course as
{\it minima} of effective actions for the spatial volume, which differ by a sign flip from the
corresponding minisuperspace actions.

This simple relation will no longer hold when other modes of the metric are included in the effective
action, because the presence of a stable ground state in the extended phase of CDT quantum gravity
indicates that there are no instabilities due to kinetic terms with the ``wrong", negative sign. 
Our current toroidal set-up has of course been designed to study effective dynamics ``beyond the scale factor"
and, more specifically, to determine a combined effective action of the scale factor and the global traceless
degrees of the metric, the moduli $\tau_i$. This is the subject of a forthcoming publication \cite{toappear}.

\section{Summary and conclusions}\label{sec:conclusions}

Motivated by the search for a larger class of observables in quantum gravity, we have initiated an investigation
into three-dimensional CDT quantum gravity for universes whose spatial topology is that of a torus.
Even before starting to analyze the dynamics of the global shape variables characterizing the torus geometries,
we were led to consider a number of interesting issues concerning the dynamics of the volume variable only.
This was triggered by the observation that typical CDT configurations on the torus with compactified time direction 
appear not to develop a nontrivial time dependence. In other words, unlike what happens 
for spherical slices, the spacetimes do not condense around a ``centre of volume", to create an extended universe 
on part of the time axis and a ``stalk" of minimal volume everywhere else.\footnote{This implies that 
in the spherical case the CDT ground state becomes essentially independent of the total time
extension $T$ as soon as $T$ becomes longer than the duration of the emergent universe. This does not seem
to happen in the toroidal case.} The reason why this happens is ultimately 
not understood and deserves further study.

We then explored the idea of forcing the system into a nontrivial time dependence by abandoning periodic
boundary conditions in time and instead using a novel type of boundary geometries, where the spatial tori 
degenerate into one-dimensional circles of a given length. 
Invoking a minisuperspace picture, we searched for a set of classical gravitational solutions in the continuum,
with the same type of boundary conditions and depending on the same number of free variables as the CDT
system, to serve as a reference point for identifying certain classical features of the continuum limit of the latter.
This led us to a generalized minisuperspace model, where (in ADM language) the Hamiltonian constraint
is not required to vanish. We found that measurements of the CDT volume profiles for a wide range of 
boundary lengths and values of the inverse Newton coupling $k_0$ can be matched with good quality
to solutions derived in this generalized mini\-superspace model. However, with the wide range of shapes 
available as classical volume profiles, this is perhaps not too surprising. Also, an asymmetric choice of boundary 
conditions in the simulations did not yield the result expected from the minisuperspace comparison.
At this point, it is difficult to disentangle whether this is due to insufficient system size or insufficient ``classicality" 
(large quantum fluctuations), or whether perhaps our identification of discrete proper time and boundary lengths 
with the corresponding continuum quantities may have been too na\"ive. 
The analysis of the dynamics of the moduli \cite{toappear} will shed some light on the size of quantum
fluctuations and the degree to which scale and shape are dynamically coupled.

As we have already commented in Sec.\ \ref{sec:introductioncdttorus}, the general issue of boundary conditions 
of the nonperturbative path integral, their relation with a continuum Hilbert space and ultimately classical interpretation 
is highly nontrivial. Comparing with 1+1 dimensional CDT quantum gravity, where the corresponding Hilbert space 
construction is under complete analytical control \cite{2dhilbert}, the situation in higher dimension is much more involved. 
Despite our choice of particularly simple ``collapsed" boundary conditions in 2+1 dimensions, we saw in
Sec.\ \ref{sec:volumeprofiles} that subtleties remain. We observed a clear effect of the boundary 
conditions on the bulk behaviour of the volume variable, but to isolate the influence of the boundary from other
effects would require a more systematic analysis, where bulk and boundary are scaled simultaneously at a specified
relative rate, similar to what can be done in CDT in 1+1 dimensions to obtain a spacetime of anti-de Sitter type
\cite{adscdt}. This is an interesting issue, but beyond the scope of this paper.

With the minisuperspace comparison remaining at this stage somewhat inconclusive, we turned to a more direct
determination of the effective action for the spatial volume from the simulation data. Here, our method of extracting it from
inverting the measured matrix of volume-volume correlation worked beautifully, and gave an unambiguous
result of the expected form for the kinetic term, which coincides with that found in the spherical case. -- This
concludes the first part of our investigation into the nonperturbative CDT description of 2+1 dimensional quantum
gravity on a torus. A natural next step will be a similar analysis of the dynamics of the moduli parameters of the same system,
involving a new set of observables relating to the global shape of the quantum universe.

\vspace{1cm}

\noindent{\bf Acknowledgments.} 
This work was supported by the Netherlands Organisation for Scientific Research (NWO) under their VICI
program. T.G. Budd acknow\-ledges support through the ERC-Advanced Grant 291092,
``Exploring the Quantum Universe'' (EQU). 
Research at Perimeter Institute is supported by the Government of Canada through Industry Canada and by 
the Province of Ontario through the Ministry of Economic Development and Innovation.
Both authors acknowledge support from Utrecht University, where part of this work was done.

\end{document}